\documentclass[aps,twocolumn,floatfix,preprintnumbers,nofootinbib]{revtex4-2}
\usepackage[utf8]{inputenc}  
\usepackage{amsmath}
\usepackage{amssymb} 
\usepackage{multirow}

\usepackage{enumerate}
\usepackage{amsmath}  
\usepackage{tikz}
\usepackage{tipa}
\usepackage{CJKutf8}
\usepackage{feynmf}
\usepackage{slashed}
\usepackage{braket}
\usepackage[toc,page]{appendix}
\usepackage{url}
\usepackage{natbib}
\usepackage{graphicx}
\usepackage[pdftex,bookmarks,linktocpage,pdfpagelabels,plainpages=false,hyperfigures,linkcolor=blue,citecolor=blue]{hyperref} %for hypertext links
\hypersetup{colorlinks=true}

\usepackage{mathrsfs,amssymb}  
\usepackage{cancel}
\usepackage[normalem]{ulem}
\usepackage{array}
\usepackage{booktabs}
\usepackage{verbatim}

\begin{document}
\begin{flushright}
MI-HET-805
\end{flushright}

\author{Bhaskar Dutta}
\email{dutta@tamu.edu}
\affiliation{Mitchell Institute for Fundamental Physics and Astronomy, Department of Physics and Astronomy, Texas A\&M University, College Station, TX 77843, USA}

\author{Doojin Kim}
\email{doojin.kim@usd.edu}
\affiliation{Department of Physics, University of South Dakota, Vermillion, SD 57069, USA}
\affiliation{Mitchell Institute for Fundamental Physics and Astronomy, Department of Physics and Astronomy, Texas A\&M University, College Station, TX 77843, USA}

\author{Hyunyong Kim}
\email{hyunyong.kim@cern.ch}
\affiliation{Mitchell Institute for Fundamental Physics and Astronomy, Department of Physics and Astronomy, Texas A\&M University, College Station, TX 77843, USA}
%%%%%%%%%%%%%%%%%%%%%%%%%%%%

\title{Proposal to use LHC general-purpose detectors in \\ ``beam-dump'' measurements for long-lived particles}

\begin{abstract} 
We propose a novel scheme for performing a beam-dump-like experiment with the general-purpose detectors (ATLAS and CMS) at the LHC.   
Collisions of high-energy protons result in jets containing a number of energetic hadrons and electromagnetic objects that are essentially ``dumped'' to hadronic and electromagnetic calorimeters, respectively, and induce the production of secondary hadrons, electrons, and photons in calorimetric showers.
We envision a situation where new physics particles are produced by the interactions of these secondary particles inside the calorimeters. 
For proof of principles, we consider the axion-like particles (ALPs) produced via the Primakoff process in the presence of their interaction with photons at CMS. 
We argue that the drift tube chambers and the ME0 module of the muon system can serve as detectors to record the photons from the ALP decay, demonstrating that assuming the background level can be controlled as discussed in this work, the resulting sensitivity reach is competitive due to their close proximity to the signal source points. 
We further show that the LHC does not suffer from a barrier, dubbed beam-dump ``ceiling'', that typical beam-dump experiments hardly surpass. This gives the LHC great potential to explore a wide range of parameter space.
This analysis can be extended to investigate various types of light mediators with couplings to the Standard Model leptons and quarks.  
\end{abstract}

\maketitle

\noindent {\bf Introduction.} 
While the existence of dark matter in the universe is clearly indicative of new physics beyond the Standard Model (SM), the experimental effort in the search for dark-matter candidates, especially, GeV-scale weakly interacting massive particles, via their hypothetical non-gravitational interactions is not yet fruitful.  
As an alternative possibility, MeV-scale dark-matter candidates are receiving great attention, as they can be thermally produced as well and are less constrained by the existing searches. 
For them to reproduce the observed dark-matter relic abundance, their interaction strengths to the SM are likely to be feeble and MeV-scale mediators often come into play. 
Therefore, the dark-matter search program is now promoted to a more generic dark-sector search program including the mediators and their feebly interacting nature motivates the intensity-frontier facilities such as beam-based neutrino experiments. 

By contrast, the general-purpose detectors (ATLAS and CMS) of the LHC, an energy-frontier facility, has been designed to be optimally sensitive to GeV-to-TeV-scale physics. 
The LHC is (indirectly) capable of probing MeV-scale physics by placing additional forward-physics facilities distant from the primary interaction point (e.g., FASER~\cite{Feng:2018pew,FASER:2018bac} and SND~\cite{SNDLHC:2022ihg}). 
On the contrary, the exploration of MeV-scale physics in the central region is still seemingly unpromising and irrelevant. 

We point out that the hadronic calorimeters (HCAL) and electromagnetic calorimeters (ECAL) of the detectors can be considered as dumps of particles inside a jet produced by a proton collision, sourcing MeV-scale new physics particles as in typical beam-dump experiments. 
We then infer the production of a new physics particle when it decays to SM particles inside the muon system. 
The large production cross-section of jets ensures the copious production of feebly-interacting new physics particles and the large angular coverage (i.e., almost $4\pi$ coverage in solid angle) of the muon system allows for a large fiducial signal flux. 
Furthermore, high-energy beam protons allow new physics particles to be energetic, hence significantly boosted, and the muon system is extremely close to the calorimeters. Therefore, it is possible to access the region of parameter space toward larger couplings and larger masses (henceforth denoted by ``prompt-decay'' regime\footnote{where the rest-frame mean decay length is much less than 1 meter.}) that typical beam-dump experiments would hardly reach. We note that searching for new particles by ``dumping'' the particles produced in the LHC collisions to LHC infrastructure~\cite{Feng:2018pew,FASER:2018bac,SNDLHC:2022ihg} or detector modules~\cite{Galon:2019owl} has also been proposed.

We illustrate the main idea with an axion-like particle (ALP) interacting with the SM photon (e.g., BC9 in Ref.~\cite{Beacham:2019nyx}):
\begin{equation}
    -\mathcal{L}_{\rm int} \supset \frac{1}{4} g_{a\gamma\gamma} a F_{\mu\nu}\tilde{F}^{\mu\nu}\,,
\end{equation}
where $a$ and $F_{\mu\nu}$ ($\tilde{F}_{\mu\nu}$) denote the ALP field and the (dual) field strength tensor of the SM photon and where $g_{a\gamma\gamma}$ parameterizes the interaction strength between the ALP and the SM photon.  
We further assume that an ALP decays to a pair of photons that leave experimental signatures at the CMS muon system. 

\medskip

\noindent {\bf Proposal outline.} The proposed measurement begins with the production of jets at the interaction point. 
A number of hadrons and electromagnetic particles inside a jet are essentially ``dumped'' and absorbed to the HCAL and ECAL, respectively. 
They eventually end up with electromagnetic showers creating photons (plus electrons and positrons) copiously, while potentially inducing the production of secondary hadrons in the midst of absorption. 
These photons can undergo the Primakoff process and convert to ALPs in the presence of a non-zero $g_{a\gamma\gamma}$. 
The produced ALPs then enter a drift tube (DT) chamber~\cite{CMS:1997dma,Hebbeker:2017bix} in the muon system and decay to photons therein. 
We also consider ME0, a projected new station in the muon system available at Run4~\cite{Hebbeker:2017bix}, as it may readily allow for dedicated triggers for signal detection.
We will elaborate on the production of ALP events, the detection principle at DT chambers and ME0, potential backgrounds, our sensitivity study, and the trigger aspects in the next sections. 

\medskip

\noindent {\bf Signal production.} 
A precise estimate of the photon flux inside the HCAL and ECAL is a key factor for a precise estimate of the sensitivity reach. 
In our study, we first utilize \texttt{Pythia8.2}~\cite{Sjostrand:2014zea} to simulate jet production. Long-lived hadrons (e.g., $\pi^\pm$ and $K^\pm$) and stable particles (e.g., $e^\pm$ and $\gamma$) are fed, respectively, into the simplified HCAL and ECAL modules built under the \texttt{GEANT4}~\cite{GEANT4:2002zbu} code package with the \texttt{QGSP\_BERT} physics list which simulates their subsequent interactions including electromagnetic showering. 
We take simplified specifications that the HCAL and ECAL modules are mostly Cu-based and PbWO$_4$-based, respectively, and their effective average lengths are set to be 1.3~m and 0.6~m, respectively.\footnote{The effective lengths covered by these materials are somewhat smaller, but the photons affecting the sensitivity reaches are mostly produced at earlier electromagnetic shower stages and our conclusions do not change with true effective lengths. Also, their dimensions transverse to the direction of the injected particles are set to be large enough as their detailed scale does not affect the photon flux substantially.}

\begin{figure}[t]
    \centering
    \includegraphics[width=0.235\textwidth]{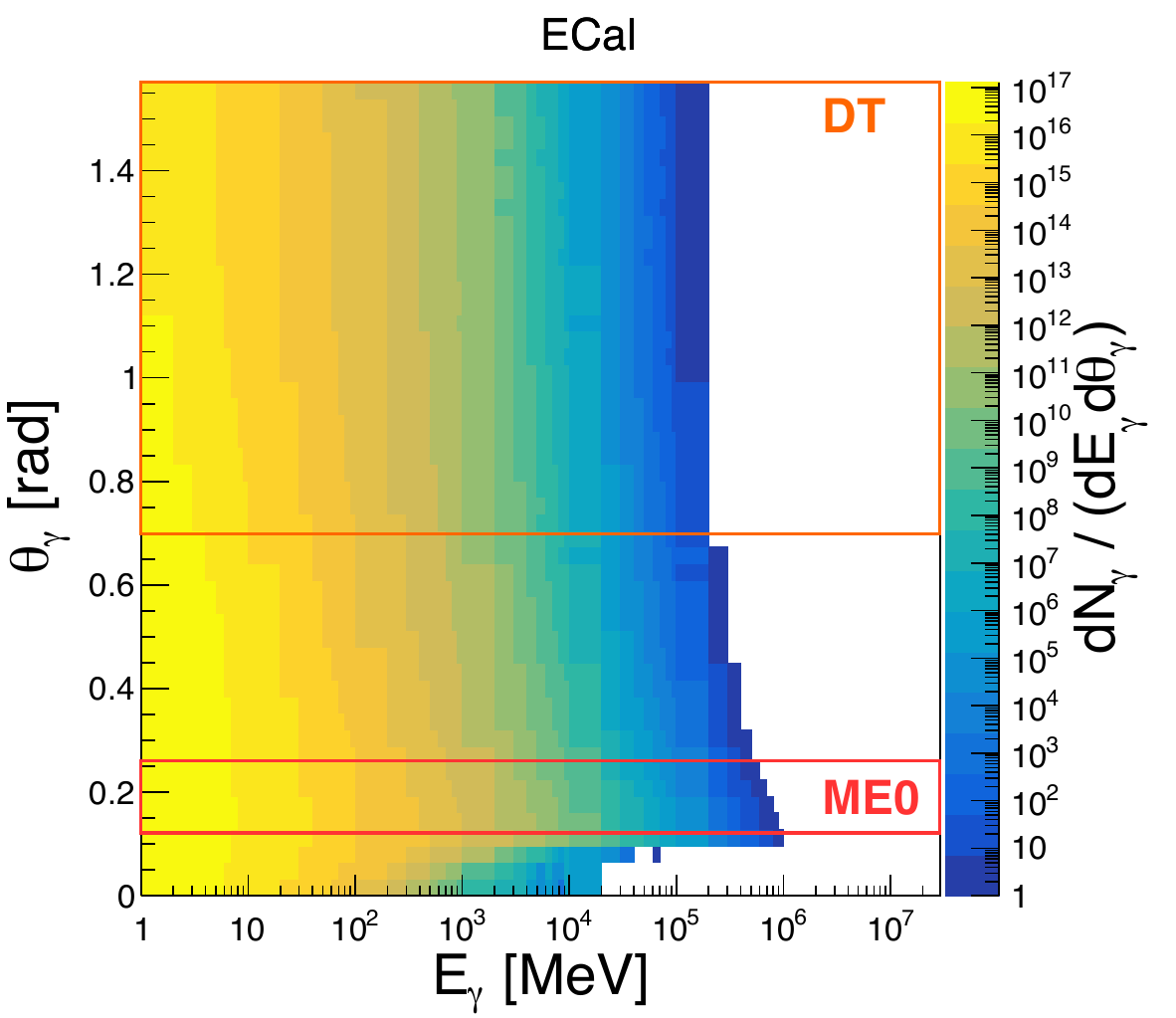}
    \includegraphics[width=0.235\textwidth]{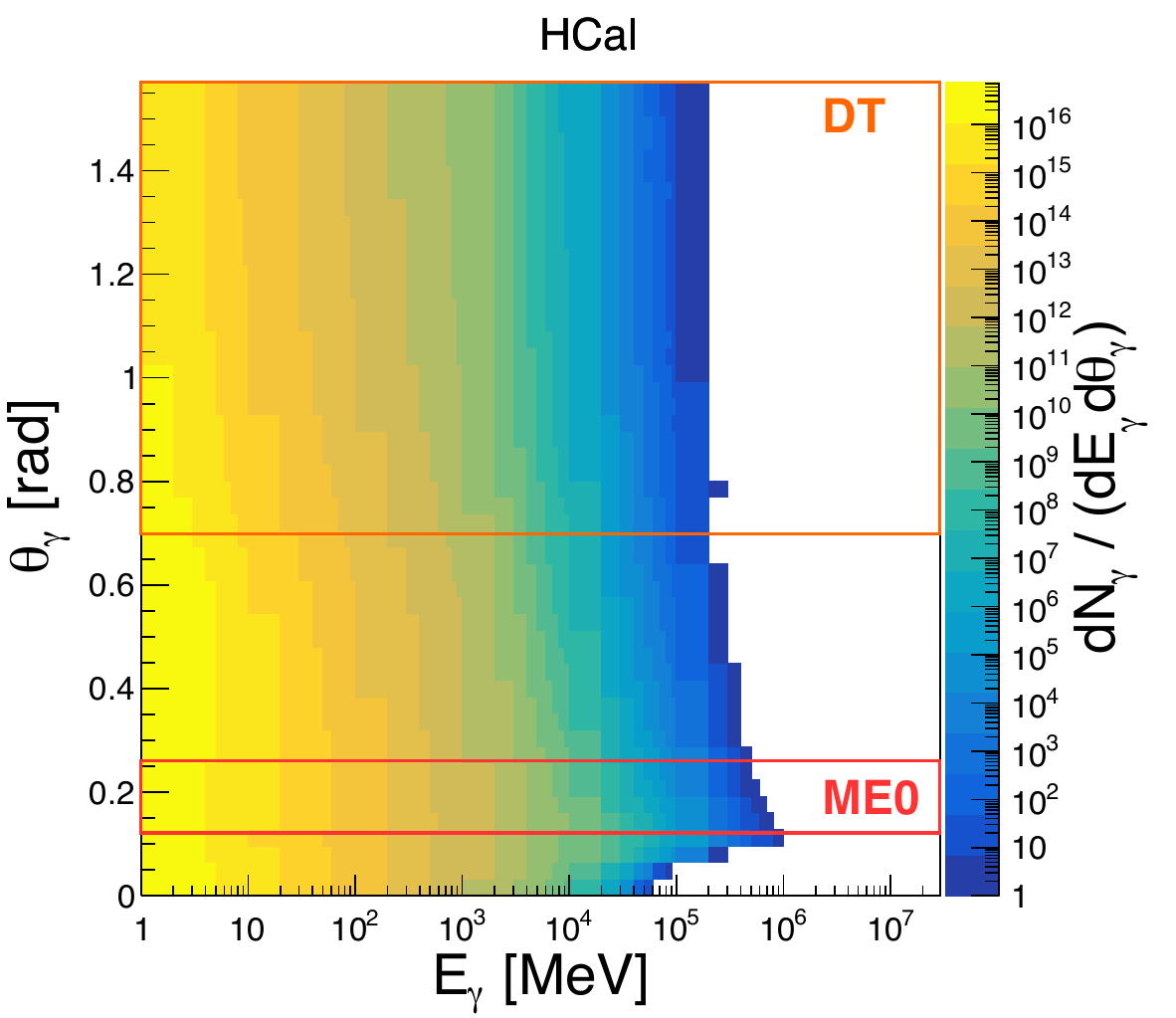}
    \caption{Energy-angle ($E_\gamma-\theta_\gamma$) correlations of the photons inside the ECAL (left) and HCAL (right) satisfying $E_\gamma>1$~MeV and $\Delta R_\gamma<0.5$. The number of photons is normalized to an integrated luminosity of 500~fb$^{-1}$ at $\sqrt{s}=13.6$~TeV. The orange and red lines delimit the angular coverages of DT chambers and ME0, respectively. See the text for more details.}
    \label{fig:photon}
\end{figure}

Denoting the photon energy by $E_\gamma$ and the angle of a photon with respect to the beam direction by $\theta_\gamma$, we show the $E_\gamma-\theta_\gamma$ correlations of the photons inside the ECAL (left) and HCAL (right) that are predicted by our \texttt{GEANT4} simulations in FIG.~\ref{fig:photon}. 
For these plots, we choose the photons satisfying $E_\gamma>1$~MeV and $\Delta R_\gamma = \sqrt{(\Delta\theta_{\gamma j})^2+(\Delta\phi_{\gamma j})^2}<0.5$ with $\Delta\theta_{\gamma j}$ and $\Delta\phi_{\gamma j}$ being the polar and azimuthal angle distances between a given photon and the center of the associated jet. 
Indeed, in regard to the data analysis, by restricting photons to within the jet radius, we isolate signals from potential backgrounds arising randomly outside the jet cone.
The number of photons is normalized to an integrated luminosity of 500~fb$^{-1}$ at $\sqrt{s}=13.6$~TeV and the total numbers of photons are  $1.5\times 10^{19}$ and $6.5\times 10^{18}$ for the ECAL and HCAL, respectively. 
While the majority of photons are soft ($\lesssim10$~MeV) and forward-moving, a sizable fraction of photons can be found in the central region.  
For example, 3.7\% (ECAL) and 3.5\% (HCAL) of photons in FIG.~\ref{fig:photon} are potentially relevant to producing signal ALPs of 100 MeV.
Also, we observe that the forward-moving photons have a tendency to be harder than the central-region photons.

As mentioned earlier, a produced photon can convert to an ALP via the Primakoff process with an atomic target (denoted by $N$) inside the calorimeters, i.e., $\gamma + N \to a + N$, and the differential production cross-section $\sigma_P$ in $\theta_a$, angle of the outgoing ALP with respect to the incoming photon is given by
\begin{equation}
    \frac{d\sigma_P}{d\theta_a}=\frac{1}{4}g_{a\gamma\gamma}^2\alpha_{\rm em} Z^2 \left|F_{\rm Helm}(t) \right|^2 \frac{p_a^4 \sin^3\theta_a}{t^2}\,
\end{equation}
where $\alpha_{\rm em}$, $Z$, and $p_a$ are the electromagnetic fine structure constant, the atomic number of the target material,
and the momentum of the outgoing ALP, and momentum transfer $t=m_a^2 - 2E_\gamma(E_a - p_a \cos\theta_a)$. 
In the collinear limit where the momentum transfer to the atomic system is negligible, $E_a \approx E_\gamma$ and $\theta_a \to 0$. We take this limit in our data analysis, i.e., the produced ALPs inherit the energy and momentum direction from the incoming photons. 
Therefore, one can straightforwardly infer the $E_a-\theta_a$ correlations for any given $m_a$ from FIG.~\ref{fig:photon}.
Finally, $F_{\rm Helm}$ describes the usual Helm form factor,
\begin{equation}
    F_{\rm Helm}(t)=\frac{3j_1(\sqrt{|t|}R_1)}{\sqrt{|t|}R_1}\exp\left( -\frac{|t|s^2}{2}\right),
\end{equation}
where $j_1$ is the spherical Bessel function, $s=0.9$~fm, and $R_1=
\sqrt{(1.23A^{1/3}- 0.6)^2 + 2.18}$~fm with $A$ being the atomic mass number of the target material~\cite{Lewin:1995rx}.

Basically, a photon may go through either the Primakoff process or the other SM processes (e.g., pair production and photoelectric absorption). Therefore, the ALP production probability $P_{\rm prod}$ is given by
\begin{equation}
    P_{\rm prod}=\frac{\sigma_P}{\sigma_{\rm SM} +\sigma_P} \approx \frac{\sigma_P}{\sigma_{\rm SM}},
\end{equation}
where we use the total cross-section of SM interactions $\sigma_{\rm SM}$ reported in Ref.~\cite{xcom} and the approximation holds for $\sigma_{\rm SM} \gg \sigma_P$. 

\medskip

\noindent {\bf Detection principle.}
Once an ALP is produced, it should first travel to a detector module of interest and decay therein.
The detection probability, say $P_{\rm det}$, is essentially governed by the exponential decay law and we have
\begin{equation}
    P_{\rm det}=\exp\left(-\frac{L}{\tilde{l}} \right)\left[1-\exp\left(-\frac{\Delta L}{\tilde{l}} \right) \right], \label{eq:Pdecay}
\end{equation}
where $L$ is the distance between the ALP production point and the detector of interest and $\Delta L$ is the travel length of the ALP within the detector of interest. 
Here $\tilde{l}$ denotes the mean decay length that can be calculated with the ALP boost factor $\gamma_a[=(1-\beta_a^2)^{-1/2}]$ and the ALP decay width $\Gamma_a$: $\tilde{l}=\beta_a c \cdot \gamma_a / \Gamma_a$ with $\Gamma_a$ given by
\begin{equation}
    \Gamma_a=\frac{1}{64\pi}g_{a\gamma\gamma}^2m_a^3\,.
\end{equation}

As mentioned earlier, we consider two types of detector modules in the muon system for our sensitivity estimates:
\begin{itemize}  \itemsep1pt \parskip0pt \parsep0pt
    \item DT chambers~\cite{CMS:1997dma,Hebbeker:2017bix}: DT chambers are located in $|\eta|<1.2$ of the barrel region (see the orange lines in FIG.~\ref{fig:photon}\footnote{Note that they show the average angular coverage as DT chambers in different stations cover slightly different angular ranges.}) and $\sim 1.2 - 1.8$ m away from the HCAL while the NbTi-based superconductor solenoid magnet is placed in-between the HCAL and DT chambers. Each of them consists of $8-12$ layers each of which is further segmented into long aluminum drift cells filled with gas. Between neighboring stage DT chambers, a few tens of cm-scale iron yokes are placed.
    \item ME0~\cite{Hebbeker:2017bix}: ME0 will be introduced to the muon system in the endcap region of CMS to maintain its performance at the high-luminosity LHC. It has a 6-layer structure and is based on the Gas Electron Multiplier technology and is intended to cover the far-forward region of $2.0<|\eta|<2.8$ (see the red lines in FIG.~\ref{fig:photon}) at the immediate downstream of the HCAL in the endcap region.
\end{itemize}
Both modules are gas-based. Basically, gaseous detectors use the ionization, drift, and diffusion processes to amplify signals generated by charged particles passing through the gas volume. 

The chamber system can detect photons, although it is primarily designed for detecting muons.
When an ALP enters a DT chamber, it can decay into two photons inside a drift cell. 
Each of them then splits into an electron-positron pair knocking electrons off the atoms of the gas. If such an electron (or positron) is energetic enough, it can pass through several layers and fire multiple drift cells, followed by getting absorbed into an iron yoke without firing the drift cells inside the next DT chamber.
Therefore, the expected experimental signature is successively fired multiple drift cells belonging to a series of layers, which collectively show a diphoton pattern aligning with the associated jet.  
Simulating signal events to see if they leave a diphoton-like pattern, firing multiple cells, requires a dedicated detector-level study, hence is beyond the scope of this paper.
We instead perform photon detection simulations with \texttt{GEANT4}~\cite{GEANT4:2002zbu}, taking a simplified geometry of multi-layered aluminum drift cells filled with argon gas and injecting photons. 
The efficiency of photon detection depends on the energy of the photons. 
Our simulation study suggests that more than 90\% of photons with energies ranging from a few MeV to GeV can pass through a single DT chamber and generate multiple hits in the drift cells of consecutive layers. In this sense, we assume that most ALPs with $E_a>10$~MeV can be identified as signal candidate events once they decay inside a single DT chamber in our analysis.

The basic concept of signal detection at ME0 is essentially the same as that of DT chambers. 
The photons from an ALP decay split into electron-positron pairs which subsequently ionize the atoms of the gas and eventually result in an electron avalanche recorded by some of the finely spaced readout strips. 
Again if the initial electron (or positron) is energetic enough, it can pass through several layers.  
Therefore, the expected experimental signature is a set of fired readout strips in multiple layers, collectively forming a diphoton pattern.  

\medskip

\noindent {\bf Background consideration.} The physics potential of our proposal depends on the background levels for the signals of interest. For the benchmark ALP signal, initial simplified studies suggest that the backgrounds should be at a negligible level. However, more detailed studies, including full simulations and data analysis, will be necessary to demonstrate this conclusively. Below, we present simplified arguments to qualitatively explain why we believe that the background can be controlled to a negligible level.

Given the detection principle, we identify two major sources that potentially give rise to background events~\cite{CMS-DP-2020-022}: escaping (or punch-through) hadrons and neutron gas. 
The work in Ref.~\cite{CMS-DP-2020-022} reports the collective background hits recorded in the DT chambers during the LHC Run2. Based on the measurement data with the zero-bias trigger, we estimate a rate of 0.45 hits per the integrated luminosity 1~nb$^{-1}$ after removing the prescale factor of the zero-bias trigger. As the $pp$ inelastic scattering cross-section at $\sqrt{s}=13$~TeV is $\sim 77$~mb~\cite{ATLAS:2022mgx} [cf. $\sim 4 \times 10^{16}$ events/(500~fb$^{-1}$) at Run4], their collective rate is roughly estimated to be $\sim 10^{-8}$/cm$^2$/event at DT chambers. 
This order-of-magnitude estimation suggests that the hit rate {\it per event} at the muon system, which could potentially form a local diphoton-like pattern, is rather small and manageable.

Although the HCAL is designed to capture and absorb hadrons, a small fraction of them may escape from the HCAL and continue to move into the muon system. 
However, these hadrons are typically energetic enough to fire the associated drift cells from the beginning of a given DT chamber to the end and even multiple DT chambers. 
Likewise, they fire the associated readout strips in every layer of ME0. 
Moreover, trajectories of charged punch-through hadrons like $\pi^\pm$ and $K^\pm$ are bent by the magnet inside the muon system, the resulting patterns are more charged-particle-like.
The experimental signatures of these sorts are clearly distinguished from those that typical ALP events would leave. 
Therefore, given the small escape rate and the difference in the detector signatures, we expect the escaping-hadron-induced backgrounds to be negligible. 

Neutron gas originates from radioactivated detector components and such neutrons can induce spontaneous photons in a temporally and spatially random manner that can further fire DT chambers or ME0. Since the typical energy of these photons is within $0.1-10$~MeV~\cite{CMS-DP-2020-022,Costantini:2020dqz}, it is challenging for them to fire more than a single drift cell in the detector modules of our interest. 
A neutron-induced photon of a few MeV can fire a handful of drift cells, but it is hard for it to accidentally create a diphoton pattern in association with the jet of interest.   
Since expected signal events can fire multiple drift cells and create an outward diphoton pattern, we expect that the neutron-gas-induced backgrounds can be significantly suppressed to a negligible level. 

In addition, neutral SM long-lived particles (e.g., $K_L^0$, $K_S^0$ etc) would sneak in DT chambers or ME0. They would then eject an electron from a nearby atom through scattering or leave a scattering or decay, e.g., $K_L^0 \to \pi^\pm \ell^\mp \nu_\ell$ with $\ell =e,\mu$, $K_S^0 \to \pi^+\pi^-$, $K_S^0 \to \pi^0\pi^0$, etc, signature. However, the trajectories of the charged particles in the final state are affected by the magnetic field inside the muon system. 
Therefore, the resulting signatures differ from diphoton-like ones.
In the case of $K_S^0 \to \pi^0\pi^0$, two diphoton patterns are expected, so they would not be mistagged as a signal unless the ``sneak-in'' rate of $K_S^0$ is large.
Since the rest-frame lifetime of $K_S^0$ is $9.0 \times 10^{-11}$~s, soft $K_S^0$ would decay earlier before reaching the muon system; $K_S^0$ with a few tens to a few hundreds of boost factor can possibly reach it after traversing the CMS solenoid. 
We conduct a \texttt{GEANT4} simulation with a simplified HCAL setup and long-lived hardrons described earlier. We observe that all energetic hadrons quickly go through secondary and tertiary interactions, resulting in the production of lower energy hadrons. More quantitatively, our simulation suggests that about 900 $K_S^0$ (including those produced through secondary and tertiary interactions of energetic hadrons) with $\sim 10~{\rm GeV}<E_{K_S^0}<50~{\rm GeV}$ would come out of the HCAL modules and reach the muon system for an integrated luminosity of 500~fb$^{-1}$. Given the $\sim 30\%$ branching fraction of $K_S^0 \to \pi^0\pi^0$, $\sim0.4$ of average decay probability within the muon system of such $K_S^0$, and a small fake chance ($\sim 5$\%, i.e., the chance of missing two photons out of four) inferred from our photon detection efficiency simulation, we expect that the number of accidental backgrounds would be of order a few or less.
 
In conclusion, considering all these aspects, we expect that negligible backgrounds are achievable upon dedicated background studies which are beyond the scope of this paper.

\medskip

\noindent {\bf Beam-dump ``ceiling'' and the LHC potential.} Before presenting our main results, we discuss the general features of sensitivity reaches in the beam-dump-type experiments and the great potential of the LHC, especially focusing on the prompt-decay regime. 
Typically in this regime, $L, \Delta L \gg \tilde{l}$ and Eq.~\eqref{eq:Pdecay} is approximated to $P_{\rm det} \approx \exp(-L/\tilde{l})$.
Suppose that $N_{\gamma}$ photons set the sensitivity at $g_{a\gamma\gamma}$ with $N_{\rm sig}$ ALP events.\footnote{For example, $N_{\rm sig}=2.3$ is for the 90\% C.L. sensitivity under a zero-background assumption.} 
If $x$-times larger data acquisition were expected, the new reach $g_{a\gamma\gamma}^\prime$ under a zero-background assumption would be
\begin{equation}
    g_{a\gamma\gamma}^\prime\approx g_{a\gamma\gamma}\sqrt{1+\frac{\log x}{\log(N_\gamma \langle P_{\rm  prod} \rangle /N_{\rm sig})-1}}\,, \label{eq:garr}
\end{equation}
where $\langle P_{\rm prod} \rangle$ denotes the average value of $P_{\rm prod}$ at $g_{a\gamma\gamma}$; $N_\gamma \langle P_{\rm  prod} \rangle \gg N_{\rm sig}$ in typical situations. 
For a sufficiently large background, $\log x$ is replaced with $\log x^{1/2}$. 
This implies that the sensitivity is effectively no longer improved even with much larger statistics due to the logarithmic behavior shown above.\footnote{This also implies that the signal production rate per injection photon barely depends on $x$ as $g_{a\gamma\gamma}$ defining the sensitivity does not change much. Due to this observation, we ignored the dependence of the ALP production rate on $g_{a\gamma\gamma}$ when deriving Eq.~\eqref{eq:garr}.}
Therefore, most of the beam-dump experiments encounter this ``ceiling'' in probing the prompt-decay regime. 
This argument is generically relevant to decay channels of long-lived particles (e.g., dark photon,  Higgs-portal scalar, etc).
For completeness purposes, we also discuss the case of $L, \Delta L \ll \tilde{l}$ in the Appendix.

However, this does {\it not} hold for the proposed beam-dump measurements at the LHC. The reason is that as we will show shortly, for the $(m_a,g_{a\gamma\gamma})$ pairs defining the sensitivity lines in the prompt-decay regime, the $L\gg \tilde{l} (\propto \gamma_a) $ condition is hardly satisfied; $L\sim \mathcal{O}(1-2\,{\rm m})$ and typical $\gamma_a$ is very large under the LHC environment.  
Therefore, we expect that our proposal does not suffer from the aforementioned beam-dump ``ceiling'' and possesses great potential for exploring a wide range of the prompt-decay regime in increasing statistics.
This essentially motivates the development of dedicated trigger algorithms to maximize signal statistics. 

\medskip

\noindent {\bf Results.} We are now in the position to report our sensitivity estimates. Four different scenarios are considered here: i) Run2 (150 fb$^{-1}$) with the zero-bias trigger~\cite{CMS:2007myk}, ii) Run2+3 ($150+250$ fb$^{-1}$) with all high-level trigger~\cite{CMS:2016ngn} adapted by the actual trigger rates recorded during Run2, iii) Run4 (500 fb$^{-1}$) with a dedicated trigger, and iv) Run4 (500 fb$^{-1}$) at ME0 with a dedicated trigger. 
Integrated luminosities of 250 fb$^{-1}$ during Run3 and 500 fb$^{-1}$ during Run4 are expected in 2025 and 2030, respectively~\cite{schedule}.
In i)-iii), DT chambers are the main detection modules while in iv) only ME0 is considered. 
Regarding the trigger, in i), we conservatively consider events collected by the existing zero-bias trigger with a prescale factor of $1.1\times 10^{-6}$~\cite{CMS:2007myk}.\footnote{Practically, we choose one of the smallest among the prescale factors applied for data collection periods during Run2.}
In ii), we consider all high-level triggers most of which involve energetic jets and/or electromagnetic objects. Our \texttt{GEANT} study suggests that the spectral behavior of shower photons is similar modulo the overall normalization, so we assume an effective zero-bias trigger with a prescale factor enhanced by an order of magnitude that corresponds to the total high-level trigger rate 1~kHz.   
Finally, in iii) and iv), we assume that dedicated trigger algorithms are developed to maximize the acceptance of the signal flux. 
For example, we envision machine-learning-based level-1 or level-2 triggers which are being actively developed for various channels~\cite{Zabi:2020gjd,trigger} and expect to adapt such techniques for our ``featureful'' signals (i.e., diphoton pattern with multiple drift cells fired) in the DT chambers and ME0 module. In particular, ME0 is a small module so it may be more feasible to implement ME0-based dedicated triggers with the total trigger rate insignificantly affected~\cite{private}.

For all scenarios, we further restrict ourselves to ALPs converted from photons whose energy is greater than 10~MeV to ensure hits in multiple layers of the detection modules (i.e., DT chambers and ME0) as mentioned earlier. 
Regarding the photon detection efficiency, our analysis assumes more than 90\%, which is based on the \texttt{GEANT} study discussed earlier. Again it assumes simplified detection modules for simulational ease but capturing essential features such as structural details and gas properties. We observe that our final results are not appreciably affected by moderate degradation (i.e., a few factors) of detection efficiency.

\begin{figure}[t]
    \centering
    \includegraphics[width=0.49\textwidth]{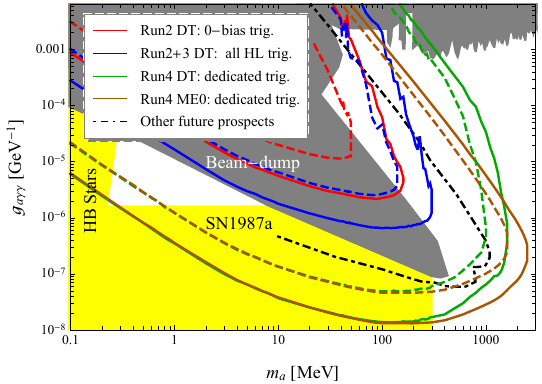}
    \caption{Expected 90\% C.L. sensitivity reaches of the ALP
decay signal at the detector modules of interest at CMS, under the assumption of negligible backgrounds (solid lines) and 100 signal events (dashed lines).
Existing laboratory-based limits~\cite{OPAL:2000puu,ALEPH:2002doz,L3:2003yon,DELPHI:2003dlq,Proceedings:2012ulb,Belle-II:2020jti,CHARM:1985anb,Bjorken:1988as,Riordan:1987aw,OPAL:2002vhf,NA64:2020qwq,Blumlein:1990ay,Aloni:2019ruo,CCM:2021lhc} are shown in gray, while astrophysics-based limits~\cite{Raffelt:1985nk,Raffelt:1987yu,Raffelt:2006cw,Payez:2014xsa,Jaeckel:2017tud} are shown in yellow. 
Future prospects compiled in Ref.~\cite{Fortin:2021cog} are also shown by the black dotdashed line.
}
    \label{fig:result}
\end{figure}

The 90\% C.L. sensitivity reaches of the above-listed four scenarios are displayed in FIG.~\ref{fig:result} in the $(m_a, g_{a\gamma\gamma})$ plane. 
Since we expect negligible backgrounds, we require the number of signal events $N_{\rm sig}$ to be 2.3 on the basis of the expected statistical error only (solid lines). 
By contrast, the dashed lines correspond to $N_{\rm sig}=100$ as conservative estimates in the case of non-zero backgrounds. 
We also show the current constraints from various existing experiments (e.g., $e^++e^- \to \gamma + {\rm inv.}$~\cite{OPAL:2000puu,ALEPH:2002doz,L3:2003yon,DELPHI:2003dlq,Proceedings:2012ulb}, Belle-II~\cite{Belle-II:2020jti},  CCM~\cite{CCM:2021lhc}, CHARM~\cite{CHARM:1985anb}, E137~\cite{Bjorken:1988as}, E141~\cite{Riordan:1987aw}, LEP~\cite{OPAL:2002vhf}, NA64~\cite{NA64:2020qwq}, $\nu$Cal~\cite{Blumlein:1990ay}, and
PrimEx~\cite{Aloni:2019ruo}), based on the limits compiled in e.g., Refs.~\cite{Bauer:2018uxu,Lanfranchi:2020crw,Fortin:2021cog}. 
The gray-colored region is based on the laboratory-produced ALP searches, and
the boundaries are set by various lepton colliders and
beam-dump experiments. 
By contrast, the regions constrained by astrophysical considerations~\cite{Raffelt:1985nk,Raffelt:1987yu,Raffelt:2006cw,Payez:2014xsa,Jaeckel:2017tud} (e.g., HB stars, supernova, etc.) are shown in yellow.\footnote{The cosmological triangle region can be probed in various ongoing neutrino experiments~\cite{dent2020new,CCM:2021lhc}. The measurement of the explosion energy
of SN1987A can provide severe tension to this region unless the star cooling process is significantly different
from the standard picture~\cite{Caputo:2021rux}.}
For comparison purposes, we also show future prospects of ongoing/planned experiments including DUNE~\cite{Brdar:2020dpr}, FASER/FASER2~\cite{Beacham:2019nyx}, LDMX~\cite{Berlin:2018bsc}, NA64~\cite{Dusaev:2020gxi}, and SHiP~\cite{Alekhin:2015byh} by the black dotdashed line.

Our sensitivity study suggests that assuming the background can be controlled, CMS can explore the regions beyond the current beam-dump limits which none of the existing (in)direct experiments have ever probed because ALPs belonging here would decay too promptly to reach their detector system. 
Note that the detector modules of interest, i.e., DT chambers and ME0, are within just a few meters from the signal production points, i.e., HCAL and ECAL. 
Moreover, due to the high-energy nature of the LHC, a sizable fraction of high-energy photons are produced. 
The ALPs from such photons are significantly boosted and thus they can be long-lived enough to reach the closely located detector modules despite fairly large values of $m_a$ and $g_{a\gamma\gamma}$. 

It is remarkable that the existing Run2 data collected with the zero-bias trigger (red line) allows for sensitivity into the unexplored regions of parameter space. 
Adding Run3 data with all high-level triggers (blue line) further extends the regions that can be probed. 
Regarding Run4, as mentioned earlier, we assume that dedicated trigger algorithms are implemented to capture the full signal flux although designing the triggers is beyond the scope of this letter. 
As expected, larger luminosity and acceptance significantly (green line) improve the sensitivity reaches and expand explorable regions, not suffering from the beam-dump ``ceiling'' as predicted earlier. 
Finally, Run4 data collected by ME0 (brown line) allows us to explore the large-mass regime above $m_a\sim 1$~GeV. 
The ME0 module will be placed in the endcap region so that it will accept more energetic signals stemming from higher-energetic photons, as also suggested by the energy-angle correlations shown in FIG.~\ref{fig:photon}. 
We further emphasize that the smallness of the ME0 module facilitates the implementation of dedicated triggers as they are less likely to affect the total trigger rate.

\medskip

\noindent {\bf Discussion and conclusions.} 
In this letter, we proposed a paradigm-shifting idea about performing a beam-dump-like measurement at the two LHC general-purpose detectors, ATLAS and CMS. 
Under the proposed measurement scheme, the two calorimeters, HCAL and ECAL, can behave as dumps to which hadrons and electromagnetic objects inside jets produced by high-energy proton collisions are dumped.
An enormous number of secondary particles including photons are produced therein while the aforementioned jet constituents are being absorbed. 
For illustration purposes, we investigated ALP production via the Primakoff process of the secondary photons and detection prospects at the muon system of CMS. 
In this regard, we carefully assessed the feasibility of DT chambers and ME0 as signal ``detectors'', identifying the expected experimental signatures and related potential backgrounds.  
Our sensitivity study suggests that CMS is capable of probing a broad range of unexplored parameter space even with the existing Run2 data collected with the zero-bias trigger. 

Our study is predicated upon a few simple assumptions and approximations. 
While we performed a simplified \texttt{GEANT}-based simulation study to assess the possibility to utilize DT chambers and ME0 as signal detectors, a more dedicated simulation study in line with careful hardware-level consideration is highly encouraged to assess the feasibility of the proposed ``beam-dump'' measurement at the LHC and identify and estimate related backgrounds more precisely.
Along the line, we expect that signal-background identification will benefit from the machine-learning-based techniques, as it is deeply related to pattern recognition. 
A trigger is another important ingredient to enhance the signal rate to be recorded. 
Given the expected experimental signatures of ALP signals, one would trigger signal events based on diphoton-like patterns in DT chambers and ME0~\cite{private} and/or adapt the existing calorimeter $E_T^{\rm miss}$ triggers, especially for high-$p_T$ signals.
In preparation for future Run4, we strongly encourage experimental collaborations to develop dedicated trigger algorithms.  

Although CMS was considered as a benchmark detector, similar measurements can be done at ATLAS as mentioned earlier. The ECAL and HCAL of ATLAS are $\sim1$ and $\sim2$ meters (i.e., $\sim 10$ interaction lengths) thick in the radial direction, respectively~\cite{ATLAS:2008xda}. In particular, the HCAL length scale is as large as CMS HCAL (1.3 m) $+$ solenoid magnet (1 m), hence HCAL is expected to suppress a similar level of punch-through hadrons entering their muon detection modules.

Finally, we emphasize that similar ideas are applicable to other new physics scenarios such as visibly-decaying bosonic mediators. Examples include ALPs interacting with leptons and models of dark photons decaying into leptons which we plan to investigate in a forthcoming publication. 
Also, this analysis in the context of light mediators (vectors and scalars) can be extended to investigate the parameter spaces associated with various anomalies, e.g., ATOMKI~\cite{Krasznahorkay:2015iga}, $g_\mu-2$~\cite{Muong-2:2021ojo}, etc.\footnote{See also Ref.~\cite{Galon:2019owl} discussing an idea of exploring the $g_\mu-2$ parameter space using the muon fixed target experiment at ATLAS.} 
Using this new scheme, we hope that ATLAS and CMS will discover new physics in the near future. 

\medskip

\section*{Acknowledgments} 
We thank Carlo Battilana, Francesca Romana Cavallo, Karri DiPetrillo, Teruki Kamon, Jason Sang Hun Lee, and Yotam Soreq for useful discussions. 
This article was supported by the computing resources of the Global Science Experimental Data Hub Center (GSDC) at the Korea Institute of Science and Technology Information (KISTI).
The work of BD, DK, and HK is supported by the U.S.~Department of Energy Grant DE-SC0010813. 

\section*{Appendix}

If the (laboratory-frame) lifetime of a produced ALP is long enough, $L, \Delta L \ll \tilde{l}$ holds. The detection probability of the ALP traveling toward the detection module of interest is approximated to 
\begin{equation}
    P_{\rm det} \approx \frac{\Delta L}{\tilde{l}}\propto g_{a\gamma\gamma}^2.
\end{equation} 
Since the ALP production rate is also proportional to $g_{a\gamma\gamma}^2$ for any small $g_{a\gamma\gamma}$, if a $x$-times larger data collection were expected, the new reach $g_{a\gamma\gamma}^\prime$ would be approximately
\begin{eqnarray}
    g_{a\gamma\gamma}^\prime &\approx& \left\{ 
    \begin{array}{l l}
    \dfrac{g_{a\gamma\gamma}}{x^{1/4}} & \hbox{for zero backgrounds,} \\ [1em]
    \dfrac{g_{a\gamma\gamma}}{x^{1/8}} & \hbox{for large backgrounds,} 
    \end{array}\right. \\ \nonumber
\end{eqnarray}
where we determine $N_{\rm sig}$, which is required to have the sensitivity, with statistical error only.  
Therefore, unlike the case of $L, \Delta L \gg \tilde{l}$, the increase of data allows to continue the exploration of more parameter space toward the delayed-decay regime, i.e., smaller $m_a$ and $g_{a\gamma\gamma}$. 

\bibliography{ref}

\end{document}